\documentclass[a4paper,12pt]{article}
\usepackage[pctex32]{graphics}
\usepackage{amssymb,amsmath}

\begin{document}
\topmargin 0pt \oddsidemargin 0mm
\newcommand{\be}{\begin{equation}}
\newcommand{\ee}{\end{equation}}
\newcommand{\ba}{\begin{eqnarray}}
\newcommand{\ea}{\end{eqnarray}}
\newcommand{\fr}{\frac}
\renewcommand{\thefootnote}{\fnsymbol{footnote}}

\begin{titlepage}

\vspace{5mm}
\begin{center}
{\Large \bf Spin-3 quasinormal modes of BTZ black hole}

\vskip .6cm

\centerline{\large
 Yun Soo Myung$^{1,a}$, Yong-Wan Kim $^{1,b}$,
and Young-Jai Park$^{2,3,c}$}

\vskip .6cm

{$^{1}$Institute of Basic Science and School of Computer Aided
Science, \\Inje University, Gimhae 621-749, Korea \\}

{$^{2}$Department of Physics and Center for Quantum Spacetime, \\
Sogang University, Seoul 121-742, Korea}\\

{$^{3}$Department of Service Systems Management and Engineering,\\
Sogang University, Seoul 121-742, Korea}

\end{center}

\begin{center}

\underline{Abstract}
\end{center}
Using the operator approach, we obtain quasinormal modes (QNMs) of
BTZ black hole in spin-3 topologically massive gravity by solving
the first-order equation of motion with the transverse-traceless
condition. We find that these are different from those obtained when
solving the second-order differential equation for the third-rank
tensor of spin-3 field subject to suitable boundary conditions and
having the sign ambiguity of mass. However, it is  shown clearly
that two approaches to the left-moving QNMs are identical, while the
right-moving QNMs of solving the second-order equation are given by
descendants of the operator approach.

\vskip .6cm

 \vskip 0.8cm

\vspace{15pt} \baselineskip=18pt
\noindent $^a$ysmyung@inje.ac.kr \\
\noindent $^b$ywkim65@gmail.com\\
\noindent $^c$yjpark@sogang.ac.kr

\thispagestyle{empty}
\end{titlepage}

\newpage
\section{Introduction}
Recently,  higher-spin theories on AdS$_3$ have been paid much
attention because they admit a truncation  to an arbitrary maximal
spin $N$~\cite{henneaux,theisen}. Especially, the  prototype of
spin-3 model is a third-rank tensor of spin-3 field  coupled to
topologically massive gravity. The authors~\cite{CLW} have discussed
the traceless spin-3 fluctuations around AdS$_3$ spacetimes and,
found that there exists a single massive propagating mode, besides
left-moving and right-moving massless modes (gauge artifacts). Also,
a trace part of spin-3 fluctuations on AdS$_3$ spacetimes has been
studied in Ref.~\cite{BLSS}. However, such a massive trace mode has
zero energy and becomes pure gauge at the chiral point. These are
considered through extended analysis of spin-2 field in the
cosmological topologically massive gravity~\cite{LSS}.

Very recently, Datta and David~\cite{DD} have solved massive wave
equations of arbitrary integer spin fields including spin-3 fields
in the BTZ black hole background, and have obtained  their
quasinormal modes which are consistent with the location of the
poles of the corresponding two-point function in the dual
conformal field theory. This could be predicted by the
AdS$_3$/CFT$_2$ correspondence. They have considered the
second-order equation of $[\bar{\Box}-m^2+4/\ell^2]
\Phi_{\rho\mu\nu}=0$ for spin-3 fields with the ingoing modes at
horizon and Dirichlet boundary condition at infinity. However, in
this case, one confronts with sign ambiguity of mass $m$. Thus, in
order to avoid this ambiguity, one could solve the first-order
equation  of
$\epsilon_\rho^{~\alpha\beta}\bar{\nabla}_\alpha\Phi_{\beta\mu\nu}+m\Phi_{\rho\mu\nu}=0$
itself with the transverse and traceless (TT) gauge condition.

On the other
hand, it was known that the operator approach (method)~\cite{SS} is very useful
to derive the quasinormal modes of spin-2 fields in the
non-rotating BTZ black hole background in the framework of
cosmological topologically massive gravity. This method has been
applied to new massive gravity to derive their quasinormal modes
of the non-rotating BTZ black hole~\cite{MKMP}.

In this work, we obtain quasinormal modes of the non-rotating BTZ
black hole in spin-3 topologically massive gravity by directly
solving the first-order equation with the TT gauge condition in
the operator approach. This method shows clearly how to derive
quasinormal modes without sign ambiguity in mass.

\section{Perturbation analysis for spin-3 fields}

Since the spin-3 fluctuations on AdS$_{3}$ or BTZ background was
formulated in~\cite{CLW}, let us write down  the perturbation
equation for the spin-3 fields  $\Phi_{\mu\nu\lambda}=e_{\mu
ab}\bar{e}^a_\nu \bar{e}^b_\lambda$ with $e_{\mu ab}$ spin-3
connection and $\bar{e}^a_\nu$ the background dreibein as
 \be \bar{\Box}
 \Phi^{\rho\alpha\beta}+\frac1{2\mu}\epsilon^{\rho\mu\nu}\bar{\nabla}_\mu
 \bar{\Box}\Phi_\nu^{\,\,\,\alpha\beta}=0.\label{spin 3 eq}
 \ee
Similar to the perturbed equation of the (spin-2) graviton,
 \be
 (\bar{\Box}+\frac{2}{l^2})h^{\rho}_{~\sigma}+\frac{1}{\mu}\epsilon^{\rho\mu\nu}\bar{\nabla}_{\mu}(\bar{\Box}+\frac{2}{l^2})h_{\nu\sigma}=0,\label{spin2eq}
 \ee
the spin-3 fluctuation also satisfies a third-order differential
equation.

In this work, we consider the non-rotating BTZ black hole with
 the mass $M=1$ and the AdS$_3$ curvature radius $\ell=1$ in global coordinates as
\begin{equation}\label{metric}
 ds^2_{\rm BTZ} = \bar{g}_{\mu\nu}dx^\mu dx^\nu
      = -\sinh^2\!\!\rho\, d\tau^2+\cosh^2\!\!\rho\, d\phi^2+d\rho^2,
\end{equation}
where the event horizon is located at $\rho=0$, while the infinity
is at $\rho=\infty$. Here we note that
$\bar{g}_{\mu\nu}=\bar{e}^a_\mu \bar{e}^b_\nu \eta_{ab}$. In terms
of the light-cone coordinates $u/v=\tau\pm \phi$, the metric
tensor $\bar{g}_{\mu\nu}$ takes the form of
\begin{equation} \bar{g}_{\mu\nu}=
\left(
  \begin{array}{ccc}
    \frac{1}{4} & -\frac{1}{4}\cosh\!2\rho & 0 \\
      -\frac{1}{4}\cosh\!2\rho & \frac{1}{4}  & 0 \\
    0 & 0 & 1 \\
  \end{array}
\right). \label{newm}
\end{equation}
Then the metric tensor (\ref{newm}) admits the Killing vector fields
$L_k$ $(k=0,-1,1)$ for the local SL(2,R)$\times$SL(2,R) algebra as
\begin{equation}
L_0=-\partial_u,~~L_{-1/1}=e^{\mp
u}\Bigg[-\frac{\cosh\!2\rho}{\sinh\!2\rho}\partial_u-\frac{1}{\sinh\!2\rho}\partial_v
\mp \frac{1}{2} \partial_\rho\Bigg],
\end{equation}
and $\bar{L}_0$ and $\bar{L}_{-1/1}$ are obtained by substituting
$u\leftrightarrow v$. Locally, they form a basis of the SL(2,R) Lie
algebra  as
\begin{equation}
[L_0,L_{\pm 1}]=\mp L_{\pm 1},~~[L_1,L_{-1}]=2L_0.
\end{equation}

In the BTZ black hole background, the spin-3 field of
$\Phi^{\rho\mu\nu}$ determined by  a third-order differential
equation (\ref{spin 3 eq}) is totally symmetric and satisfies the TT
gauge condition
 \be
 \Phi_\mu^{~\mu\nu}=0,~~~\bar{\nabla}^\mu\Phi_{\mu\nu\rho}=0.
 \ee
 Hence its number of propagating degrees of freedom is counted to
 be one as
 \be
 10-3-6=1,
 \ee
which corresponds to a single massive propagating mode as that in
AdS$_3$ background~\cite{CLW}. The third-order equation~(\ref{spin
3 eq}) can also be expressed as
 \be\label{3rdeq2}
 ({\cal D}^M{\cal D}^L{\cal D}^R \Phi)^{\rho\mu\nu}=0
 \ee
in terms of mutually commuting operators  of
 \ba
 ({\cal D}^{L/R})^{\rho\nu}=\delta^{\rho\nu}\pm \frac{1}{2}\epsilon^{\rho\mu\nu}\bar{\nabla}_\mu,
 ~~({\cal D}^M)^{\rho\nu}=\delta^{\rho\nu}+\frac{1}{2\mu}\epsilon^{\rho\mu\nu}\bar{\nabla}_\mu.
 \ea
At the critical point of $\mu=1$, the operators ${\cal D}^M$ and
${\cal D}^L$ degenerate. We note that Eq. (\ref{3rdeq2}) is reduced
to Eq. (\ref{spin 3 eq}) when using  the BTZ background
 \ba
 \bar{R}_{\rho\sigma\mu\nu}=-(\bar{g}_{\rho\mu}\bar{g}_{\sigma\nu}-\bar{g}_{\rho\nu}\bar{g}_{\sigma\mu}),
 ~~~\bar{R}_{\mu\nu=}=-2\bar{g}_{\mu\nu},
  \ea
together with the TT gauge condition and the relation of
$[\bar{\nabla}_\mu,\bar{\nabla}_\nu]
\Phi^{\mu\alpha\beta}=-4\Phi_\nu^{~\alpha\beta}$. Therefore, the
third-order equation~(\ref{spin 3 eq}) can be decomposed into
three first-order differential equations:
 \be\label{three1st}
 ({\cal D}^M \Phi)^{\rho\mu\nu}=0,~~({\cal D}^L
 \Phi)^{\rho\mu\nu}=0,~~({\cal D}^R \Phi)^{\rho\mu\nu}=0,
 \ee
for a massive, a left-moving, and a right-moving degree of
freedom, respectively.

Importantly, three first-order differential
equations~(\ref{three1st}) can be simply rewritten in terms of a
single massive first-order differential equation as
 \be\label{meof}
 \epsilon_\rho^{~\alpha\beta}\bar{\nabla}_\alpha\Phi_{\beta\mu\nu}+m\Phi_{\rho\mu\nu}=0
 \ee
with $m=2\mu,~2$, and $-2$. On the other hand, it could also be
expressed in terms of a second-order differential
equation~\cite{DD} as
 \be \label{meos}
 \Big[\bar{\Box}^2-m^2+4\Big]\Phi_{\rho\mu\nu}=0.
 \ee
At this stage, we wish to point out the presence of sign ambiguity
$\pm m$ in the second-order equation (\ref{meos}). In order to
avoid this ambiguity, one could directly solve the first-order
equation (\ref{meof}) with the TT gauge condition.

Having the structure in mind, let us find quasinormal modes for
the spin-3 field in the BTZ background by solving the equation of
motion (\ref{meof}) with the TT gauge condition. In order to
implement the operator method~\cite{SS,MKMP}, let us choose either
the anti-chiral highest weight condition of
$L_1\Phi_{\rho\mu\nu}=0$ or the chiral highest weight condition of
$\bar{L}_1\Phi_{\rho\mu\nu}=0$, but not both simultaneously.
Actually,   we note that   for a generic symmetric tensor
$\Phi_{\rho\mu\nu}$, the transversality condition of
$\bar{\nabla}^\mu\Phi_{\mu\nu\rho}=0$ is not equivalent to
choosing the chiral (anti-chiral) highest weight condition.

\section{Left-moving quasinormal modes}

The least damped ($n=0$) quasinormal mode can be found by
considering the form
 \be\label{ansatz}
 \Phi_{\rho\mu\nu}(u,v,\rho)=e^{-i\omega \tau-ik\phi} F_{\rho\mu\nu}(\rho)
  =e^{-ihu-i\bar{h}v}F_{\rho\mu\nu}(\rho)
 \ee
with $\omega=h+\bar{h}$ and $k=h-\bar{h}$. This is the primary field
which satisfies
 \be
 L_0\Phi_{\rho\mu\nu}(u,v,\rho)=ih\Phi_{\rho\mu\nu}(u,v,\rho),
  ~~\bar{L}_0\Phi_{\rho\mu\nu}(u,v,\rho)=i\bar{h}\Phi_{\rho\mu\nu}(u,v,\rho).
 \ee
Note here that the subscript $\rho$ in
$\Phi_{\rho\mu\nu}(u,v,\rho)$ is a dummy index, while $\rho$ in
the argument is the radial coordinate in (\ref{metric}). It seems
to be a formidable task to solve the first-order equation with the
TT gauge condition without choosing a simplified form of
$F_{\rho\mu\nu}$. Inspired by the lesson learned from the spin-2
analysis~\cite{SS,MKMP}, after tedious computations,  we find
 the
explicit solution
 \ba\label{leftm1}
 F_{u\mu\nu}(\rho)&=&\left(
  \begin{array}{ccc}
    0 & 0 & 0\\
    0 & 0 & 0 \\
    0 & 0 & 0 \\
  \end{array}
 \right), \\
  \label{leftm2}
 F_{v\mu\nu}(\rho)&=&\left(
  \begin{array}{ccc}
    0 & 0 & 0\\
    0 & 1 & \frac{2}{\sinh2\rho} \\
    0 & \frac{2}{\sinh2\rho} & \frac{4}{\sinh^22\rho} \\
  \end{array}
 \right)F_{vvv}(\rho),\\
  \label{leftm3}
 F_{\rho\mu\nu}(\rho)&=&\left(
  \begin{array}{ccc}
    0 & 0 & 0\\
    0 & \frac{2}{\sinh2\rho} & \frac{4}{\sinh^22\rho}\\
    0 & \frac{4}{\sinh^22\rho} & \frac{8}{\sinh^32\rho} \\
  \end{array}
 \right)F_{vvv}(\rho),
 \ea
which imply that $F_{vvv}(\rho)$ is a single massive propagating
mode. Here Eq.~(\ref{leftm2}) is similar to the spin-2
case~\cite{SS}, while Eqs.~(\ref{leftm1}) and (\ref{leftm3})
represent new features of the spin-3 field.

Under the form of $\Phi_{\rho\mu\nu}(u,v,\rho)$ in
Eq.~(\ref{ansatz}) with $F_{\rho\mu\nu}(\rho)$ in
Eqs.~(\ref{leftm1})-(\ref{leftm3}), the transversality condition of
$\bar{\nabla}^\mu\Phi_{\mu\nu\rho}=0$ is equivalent to the
anti-chiral highest weight condition of $L_1\Phi_{\rho\mu\nu}=0$,
giving the constraint
 \be\label{constraint1}
 \sinh2\rho\Big[\frac{d}{d\rho}F_{vvv}(\rho)\Big]+2i(\bar{h}+h\cosh2\rho)F_{vvv}(\rho)=0.
 \ee
We emphasize that Eq. (\ref{constraint1}) takes the form of the
equation for the scalar field $ F_{vvv}(\rho)$, not a third-rank
tensor field. Then, its solution is given by
 \be
 F_{vvv}(\rho)=C(\sinh2\rho)^{-ih}(\tanh\rho)^{-i\bar{h}}
 \ee
with a constant $C$. Finally, the equation of motion (\ref{meof})
determines $h$ as a function of  $m$ of
 \be
 h=-ih_L(m),~~~h_L(m)=\frac{1}{2}(m-2).
 \ee
Thus, the solution is summarized as
 \be\label{leftsol01}
 \Phi^L_{\rho\mu\nu}(u,v,\rho)=e^{ik(\tau-\phi)-2h_L(m)\tau}F^L_{\rho\mu\nu}(\rho)
 \ee
where $F^L_{\rho\mu\nu}(\rho)$ is given by
Eqs.~(\ref{leftm1})-(\ref{leftm3}) with
 \be
 F^L_{vvv}(\rho)=(\sinh\rho)^{-2h_L(m)}(\tanh\rho)^{ik}.
 \ee
Considering the form of quasinormal frequency
 \be
 \omega=\omega_{\rm Re}-i\omega_{\rm Im},
 \ee
we read off it from Eq. (\ref{leftsol01})
 \be
 \omega_L=-k-2ih_L(m).
 \ee
Thus, the solution (\ref{leftsol01}) corresponds to a left-moving
massive quasinormal mode of the least damped ($n=0$) case for
$m=2\mu$, leading to
 \be
 h_L(\mu)=\mu-1>0~~\mu>1.
 \ee
As is expected by the anti-chiral gravity, we observe that there
is no quasinormal modes ($h_L=0$) at the  anti-chiral point of
$\mu=1$. Also we observe that the asymptotic $\rho$-dependence of
$F^L_{vvv}$ takes the form of $F^L_{vvv}\sim e^{2(1-\mu)\rho}$,
which is compared to the spin-2 asymptotic dependence of
$F_{vv}\sim e^{(1-\mu)\rho}$ for $m=\mu$~\cite{BMS}.

In order to derive the higher-order quasinormal modes, we act on
the anti-chiral highest weight quasinormal modes with the operator
of $\bar{L}_{-1}L_{-1}$. The effect of this will be to replace
$\omega_{\rm Im}$ by  $\omega_{\rm Im}+2$ in Eq.
(\ref{leftsol01}). Hence one could expect to have
\be\label{leftsoln}
 \Phi^{(n)L}_{\rho\mu\nu}(u,v,\rho)= \Big(\bar{L}_{-1}L_{-1}\Big)^n
 \Phi^L_{\rho\mu\nu}(u,v,\rho),
 \ee
which are descendents of $\Phi^L_{\rho\mu\nu}(u,v,\rho)$. Since
$\bar{L}_{-1}L_{-1}$ commutes with the equation (\ref{meof}),
$\Phi^{(n)L}_{\rho\mu\nu}(u,v,\rho)$ is again the solution to the
first-order equation with the same boundary condition of
asymptotic fall-off as in $\Phi^L_{\rho\mu\nu}(u,v,\rho)$. Hence,
the complete tower of the left-moving spin-3 quasinormal modes
could be generated from $\Phi^L_{\rho\mu\nu}(u,v,\rho)$.

 Consequently,
the corresponding quasinormal frequencies are given by \be \label{lmqnm}
\omega^n_L=-k-2i\Big(h_L(\mu)+n\Big), ~~n\in Z. \ee

\section{Right-moving quasinormal modes}

 On the other hand,
right-moving quasinormal modes of the least damped case can be
obtained by substitution of  $u\rightarrow v$,
$h\rightarrow\bar{h}$, and $m\rightarrow -m$. Explicitly, they take
the form of
 \ba\label{rightm1}
 F_{u\mu\nu}(\rho)&=&\left(
  \begin{array}{ccc}
    1 & 0 & \frac{2}{\sinh2\rho}\\
    0 & 0 & 0 \\
    \frac{2}{\sinh2\rho} & 0 & \frac{4}{\sinh^22\rho} \\
  \end{array}
 \right)F_{uuu}(\rho), \\
 \label{rightm2}
 F_{v\mu\nu}(\rho)&=&\left(
  \begin{array}{ccc}
    0 & 0 & 0\\
    0 & 0 & 0 \\
    0 & 0 & 0 \\
  \end{array}
 \right),\\
 \label{rightm3}
 F_{\rho\mu\nu}(\rho)&=&\left(
  \begin{array}{ccc}
    \frac{2}{\sinh2\rho} & 0 & \frac{4}{\sinh^22\rho}\\
    0 & 0 & 0\\
    \frac{4}{\sinh^22\rho} & 0 & \frac{8}{\sinh^32\rho} \\
  \end{array}
 \right)F_{uuu}(\rho),
 \ea
which imply that $F_{uuu}(\rho)$ is a single massive propagating
mode. Here Eq.~(\ref{rightm1}) is similar to the spin-2
case~\cite{SS}, while Eqs.~(\ref{rightm2}) and (\ref{rightm3})
represent new features of the spin-3 field.

The transversality condition of
$\bar{\nabla}^\mu\Phi_{\mu\nu\rho}=0$ in this case is now
compatible with the chiral highest weight condition of
$\bar{L}_1\Phi_{\rho\mu\nu}=0$, giving the differential equation
of $F_{uuu}(\rho)$
 \be\label{constraint}
 \sinh2\rho\Big[\frac{d}{d\rho}F_{uuu}(\rho)\Big]+2i(h+\bar{h}\cosh2\rho)F_{uuu}(\rho)=0.
 \ee
The solution is given by
 \be
 F_{uuu}(\rho)=D(\sinh2\rho)^{-i\bar{h}}(\tanh\rho)^{-ih}
 \ee
with a constant $D$. The equation of motion (\ref{meof}) determines
$\bar{h}$ as a function of $m$
 \be
 \bar{h}=-ih_R(m),~~~h_R(m)=\frac{1}{2}(m-2).
 \ee
 Considering $m=-2\mu$, one finds
 \be h_R(\mu)=-\mu-1\equiv \tilde{\mu}-1>0,~~\mu<-1~ (\tilde{\mu}>1).\ee
Then, the $n=0$ least damped right-moving solution is given by
 \be\label{leftsol}
 \Phi^R_{\rho\mu\nu}(u,v,\rho)=e^{-ik(\tau+\phi)-2h_R(\mu)\tau}F^R_{\rho\mu\nu}(\rho)
 \ee
where  $F^R_{\rho\mu\nu}(\rho)$ is given by
Eqs.~(\ref{rightm1})-(\ref{rightm3}) with
 \be\label{leftsol1}
 F^R_{uuu}(\rho)=(\sinh\rho)^{-2h_R(m)}(\tanh\rho)^{-ik}.
 \ee
Thus, its quasinormal mode can be read off as
 \be
 \omega_R=k-2ih_R(\mu).
 \ee
As is expected by the chiral gravity, we observe that there is no
quasinormal modes ($h_R=0$) at the  chiral point of $\mu=-1$.

Similarly, the higher-order quasinormal modes are obtained by
acting the operator of $\bar{L}_{-1}L_{-1}$ as
\be\label{rightsoln}
 \Phi^{(n)R}_{\rho\mu\nu}(u,v,\rho)= \Big(\bar{L}_{-1}L_{-1}\Big)^n
 \Phi^R_{\rho\mu\nu}(u,v,\rho),
 \ee
which are descendants of $\Phi^R_{\rho\mu\nu}(u,v,\rho)$. Its
quasinormal frequencies are given by
 \be \label{rmqnm}
 \omega_R^n=k-2i\Big(h_R(\mu)+n\Big).
 \ee

In Table~1, we have briefly summarized the results by comparing the
spin-2 field in Ref.~\cite{SS,MKMP} with the spin-3 topologically
massive gravity. For the spin-2 field satisfying (\ref{spin2eq}),
the QNMs for the left-moving component exist for only $\mu>1$, while
the QNMs for the right-moving one for only $\mu<-1$ as in
Ref.~\cite{SS,MKMP}. Especially, the result of Ref.~\cite{SS} is
obtained by $u\rightarrow v$, $h\rightarrow\bar{h}$, but not by
$m\rightarrow -m$. Instead, the authors gave the mass ranges for the
QNMs as the left-moving (right-moving) component for $\mu>1$
($\mu<-1$), which are exactly the same with replacing $m$ by $-m$ in
the equation of motion as shown in Table.~1. The QNMs for the
left-moving (right-moving)  spin-3 field are, by the same token,
valid for only $\mu>1$ $(\mu<-1)$. Here we also note that there are
no QNMs at $\mu=1$ ($\mu=-1$) for the left-moving (right-moving)
spin-2 field, while at $\mu=1$ ($\mu=-1$) for the left-moving
(right-moving) spin-3 field, expected by anti-chiral (chiral)
gravity, respectively.

\begin{table}[ht]
\label{SSours}
\centering
\begin{tabular}{c c c}
\hline\hline
\multicolumn{3}{c} {Solutions of first-order differential equations}\\
\hline
\multicolumn{3}{c} {spin-2 field}\\
\hline
 & {$\epsilon_\mu^{~\alpha\beta}\bar{\nabla}_\alpha h_{\beta\nu}+m h_{\mu\nu}=0$}
 & {$\epsilon_\mu^{~\alpha\beta}\bar{\nabla}_\alpha h_{\beta\nu}-m h_{\mu\nu}=0$}  \\
 &  $L_1 h_{\mu\nu}=0$  &   $\bar{L}_1 h_{\mu\nu}=0$  \\
 &  $h_L(m)=\left.\frac{m}{2}-\frac{1}{2}\right|_{m=\mu}=\frac{\mu}{2}-\frac{1}{2}$
           &   $h_R(m)=\left.\frac{m}{2}-\frac{1}{2}\right|_{m=-\mu}=-\frac{\mu}{2}-\frac{1}{2}$ \\
 &  $\omega^L_n=-k-2i\left(h_L(m)+n\right)$  &   $\omega^R_n=k-2i\left(h_R(m)+n\right)$  \\
\hline
\multicolumn{3}{c} {spin-3 field}\\
\hline
 & $\epsilon_\rho^{~\alpha\beta}\bar{\nabla}_\alpha\Phi_{\beta\mu\nu}+m\Phi_{\rho\mu\nu}=0$
 &  $\epsilon_\rho^{~\alpha\beta}\bar{\nabla}_\alpha\Phi_{\beta\mu\nu}-m\Phi_{\rho\mu\nu}=0$ \\
 &   $L_1 \Phi_{\rho\mu\nu}=0$  &    $\bar{L}_1 \Phi_{\rho\mu\nu}=0$  \\
 &  $h_L(m)=\left.\frac{m}{2}-1\right|_{m=2\mu}=\mu-1$
           &   $h_R(m)=\left.\frac{m}{2}-1\right|_{m=-2\mu}=-\mu-1$ \\
 &  $\omega^L_n=-k-2i\left(h_L(m)+n\right)$  &   $\omega^R_n=k-2i\left(h_R(m)+n\right)$  \\
\hline
\hline
\end{tabular}
\caption{Summary of the QNMs by comparing the spin-2 field in
Ref.~\cite{MKMP} with the spin-3 field in topologically massive
gravity: The right-moving solution is obtained by solving the
first-order equation of motion with the replacement of $u\rightarrow
v$, $h\rightarrow\bar{h}$, and $m\rightarrow -m$. The equation of
motion with $m$ allows only the left-moving (anti-chiral) solution,
while the one with $-m$ gives only the right-moving (chiral)
solution in Ref.~\cite{MKMP}.}
\end{table}

\section{Discussions}
We have obtained quasinormal modes of BTZ black hole in spin-3
topologically massive gravity by directly solving the first-order
equation with the transverse-traceless condition in the operator
approach. We have found that there is no $n=0$ quasinormal modes
($h_{L/R}=0$) at the  anti-chiral/chiral point of $\mu=\pm 1$.

It seemed that these are different from those with $T_{L/R}=\frac{1}{2\pi}$~\cite{DD}
\be
\omega^n_{sL}=k-2\pi T_L i(2n+m+1-s),~~\omega^n_{sR}=-k-2\pi T_R
i(2n+m+1+s),
\ee
which are obtained when
solving the second-order differential equations for the $s$-rank
tensor of spin-$s$ field imposed by the boundary conditions. We note
that the signs $\pm$ of real part $\omega^n_{sL/R}$ are different
from $\mp$ of $\omega^n_{L/R}$. We adhere to the convention of
Ref~.\cite{SS} for the left/right-moving modes.  Comparing the
left-moving QNMs in Eq. (\ref{lmqnm}) with $\omega^n_{3L}$ with $m=2\mu$
leads to the same expression for the imaginary sector. The same
thing happens for the spin-2 when comparing $\omega^n_{2L}$ with
$m=\mu$.

On the other hand, it seems that the right-moving QNMs in Eq.
(\ref{rmqnm}) are different from $\omega^n_{3R}$. However,
$\omega^n_{3R}$ cloud be recovered from the descendants of
$\Phi^R_{\rho\mu\nu}(u,v,\rho)$~\cite{BMS}. That is, one has that
imaginary [$\omega^n_{3R}$]=imaginary [$\omega^{n+3}_{R}$] for
spin-3. Similarly, one has that imaginary
[$\omega^n_{2R}$]=imaginary [$\omega^{n+2}_{R}$] for spin-2. We have
constructed  these descendants for the spin-2 case in Appendix A and
the spin-3 case in Appendix B, whose asymptotic forms of relevant
part are consistent with those obtained by solving the second-order
differential equations.

For $\mu=\pm 1$, we expect to develop  logarithmic modes of spin-3
field in the BTZ black hole background as did for spin-2
field~\cite{MKMP,sachs}.

 Consequently, the operator approach  combined with the first-order
equation is a useful method to derive QNMs of the BTZ black hole
in the spin-$s$ topologically massive gravity.  We  suggest   that
two approaches to the left-moving QNMs are identical, while the
right-moving QNMs of  solving second-order equation are given by
descendants of the operator approach.

\section*{Acknowledgement}
Two of us (Y. S. Myung and Y.-W. Kim) were supported by the National
Research Foundation of Korea (NRF) grant funded by the Korea
government (MEST) (No.2011-0027293). Y.-J. Park was partially
supported by the National Research Foundation of Korea (NRF) grant
funded by the Korea government (MEST) through the Center for Quantum
Spacetime (CQUeST) of Sogang University with grant number
2005-0049409, and was also supported by World Class University
program funded by the Ministry of Education, Science and Technology
through the National Research Foundation of Korea(No. R31-20002).

\section*{Appendix: Descendants of spin-$s$ field}
\subsection*{A. Descendants of spin-2 field}

It was known  that by solving the first-order equation of
$\epsilon_\mu^{~\alpha\beta}\bar{\nabla}_\alpha h_{\beta\nu}+m
h_{\mu\nu}=0$ with  the TT condition, one has the ingoing highest
 weight solution for the left-moving spin-2 field near the horizon~\cite{MKMP,BMS}
 \be
 h^L_{\mu\nu}=e^{(1-\mu)t+ik(t-\phi)}(\sinh\rho)^{1-\mu}(\tanh\rho)^{ik}
              \left(\begin{array}{ccc}
       0 & 0 & 0 \\
       0 & 1 & \frac{2}{\sinh\!2\rho} \\
       0 &  \frac{2}{\sinh\!2\rho}  & \frac{4}{\sinh^2\!2\rho}
       \end{array}
  \right).
 \ee
On the other hand, by the substitution of  $u\rightarrow v$,
$h\rightarrow\bar{h}$, and $m\rightarrow -m$, we have the ingoing
highest weight solution for the right-moving spin-2 field near the
horizon
 \be
 h^R_{\mu\nu}=e^{(\mu+1)t-ik(t+\phi)}(\sinh\rho)^{\mu+1}(\tanh\rho)^{-ik}
              \left(\begin{array}{ccc}
       1 & 0 & \frac{2}{\sinh\!2\rho} \\
       0 & 0 & 0 \\
       \frac{2}{\sinh\!2\rho} &  0 & \frac{4}{\sinh^2\!2\rho}
       \end{array}
  \right).
 \ee
By acting the operator of $\bar{L}_{-1}L_{-1}$, the second
descendent for the right-moving mode is given by
 \ba
  h^{(2)R}_{\mu\nu} &=& \Big(\bar{L}_{-1}L_{-1}\Big)^2h^R_{\mu\nu}  \nonumber\\
                    &=& \frac{1}{8}e^{-2h_R(\mu)t-ik(t+\phi)}(\sinh\rho)^{\mu-3}(\tanh\rho)^{-ik}
                      \left(\begin{array}{ccc}
       m_{uu} & 0 &  \frac{m_{u\rho}}{\sinh\!2\rho} \\
       0 & 0 & 0 \\
      \frac{m_{u\rho}}{\sinh\!2\rho} &  0 & \frac{m_{\rho\rho}}{\sinh^2\!2\rho}
       \end{array}
  \right), \nonumber\\
 \ea
where
 \be
 h_R(\mu)=-\frac{\mu}{2}+\frac{3}{2},
 \ee
and
 \ba
  m_{uu} &=& (\mu(1+\mu)-k^2-ik(1+2\mu))\times \nonumber\\
         &&  \left(16-13\mu+3\mu^2-8k^2+4\mu(-3+\mu)\cosh\!2\rho+\mu(1+\mu)\cosh\!4\rho \right.\nonumber\\
           &&~ -8ik(-3+\mu+\mu\cosh\!2\rho)),
 \ea
 \ba
 m_{u\rho}&=&\frac{1}{8\cosh^4\!\rho}
              \{ -16+40\mu+139\mu^2-150\mu^3+35\mu^4-k^2(172-475\mu+243\mu^2) \nonumber\\
         &&~~  +48k^4-ik(100+283\mu-465\mu^2+150\mu^3+16k^2(10-11\mu))\nonumber\\
       &+& 4\left(2(1-\mu)^2(-11-15\mu+7\mu^2)-k^2(39-181\mu+90\mu^2)+16k^4 \right.\nonumber\\
         &&~~+ik(7-102\mu+177\mu^2-58\mu^3-62k^2(1-\mu)))\cosh\!2\rho\nonumber\\
       &+& 4\left(-2+6\mu+21\mu^2-26\mu^3+7\mu^4-k^2(25-73\mu+35\mu^2)+4k^4 \right.\nonumber\\
         &&~~-ik(15+41\mu-75\mu^2+26\mu^3+4k^2(6-5\mu)))\cosh\!4\rho\nonumber\\
       &-& 4(1-\mu)(2(1-2\mu^2+\mu^3)+k^2(5-6\mu)-ik(1-9\mu+6\mu^2-2k^2))\cosh\!6\rho\nonumber\\
       &+& \mu(1-\mu)(\mu(1-\mu)+k^2-ik(1-2\mu))\cosh\!8\rho \},
 \ea
 \ba
 m_{\rho\rho}&=& \frac{1}{4\cosh^4\!\rho}
                 \{140-196\mu+391\mu^2-210\mu^3+35\mu^4-k^2(502-665\mu+243\mu^2) \nonumber\\
           &&~  +48k^4-ik(14+823\mu-651\mu^2+150\mu^3+16k^2(14-11\mu))\nonumber\\
       &-& 8\left(28+44\mu-77\mu^2+42\mu^3-7\mu^4+k^2(62-133\mu+45\mu^2)-8k^4 \right.\nonumber\\
         &&~~-ik(86-153\mu+129\mu^2-29\mu^3-k^2(46-31\mu)))\cosh\!2\rho\nonumber\\
       &+& 4\left(28-44\mu+75\mu^2-42\mu^3+7\mu^4-k^2(82-121\mu+35\mu^2)+4k^4 \right.\nonumber\\
         &&~~+ik(2-\mu)(1-71\mu+26\mu^2-20k^2))\cosh\!4\rho\nonumber\\
       &-& 8(2-\mu)(2+3\mu-4\mu^2+\mu^3+k^2(5-3\mu)-ik(5-9\mu+3\mu^2-k^2))\cosh\!6\rho\nonumber\\
       &+&
       (1-\mu)(2+\mu)((1-\mu)(2-\mu)-k^2+ik(3-2\mu))\cosh\!8\rho\}.
 \ea
From $(\sinh\rho)^{(\mu-3)}$ in Eq. (45), its asymptotic form is
given by
 \be
 h^{(2)R}_{\rho\rho}\sim e^{(\mu-3)\rho},~~ \rho \to\infty,
 \ee
which is consistent with that of $n=0$ right-moving quasinormal
modes obtained by solving the second-order differential
equation~\cite{DD}. Explicitly, from Eq. (3.55) in Ref.~\cite{DD},
we recover the same asymptotic form of $R_{22}(\xi)\sim
e^{(m-3)\xi}$ for $\xi=\rho$ and $m=\mu$.

 \subsection*{B. Descendants of spin-3 field}

As was done  in the spin-2 field, from the right-moving highest
weight solution for the spin-3 field in Eqs.~(\ref{leftsol}) and
(\ref{leftsol1}), the third descendent quasinormal modes  can be
computed as
 \ba
  \Phi^{(3)R}_{\rho\mu\nu} &=& \Big(\bar{L}_{-1}L_{-1}\Big)^3\Phi^R_{\rho\mu\nu}  \nonumber\\
                    &=& \frac{1}{8}e^{-2h_R(\mu)t-ik(t+\phi)}(\sinh\rho)^{2(\mu-2)}(\tanh\rho)^{-ik}
                        F^{(3)}_{\rho\mu\nu}(\rho),  \nonumber\\
 \ea
where
 \be
 h_R(\mu)=-\mu+2,
 \ee
and
 \ba\label{3rddes}
  F^{(3)}_{u\mu\nu}(\rho) &=& \left(\begin{array}{ccc}
                        m_{uuu} & 0 & \frac{m_{uu\rho}}{\sinh\!2\rho} \\
                        0 & 0 & 0 \\
                        \frac{m_{uu\rho}}{\sinh\!2\rho} &  0 &
                        \frac{m_{u\rho\rho}}{\sinh^2\!2\rho}
                              \end{array}\right), \nonumber\\
  F^{(3)}_{v\mu\nu}(\rho) &=& \left(\begin{array}{ccc}
                        0 & 0 & 0 \\
                        0 & 0 & 0 \\
                        0 & 0 & 0
                              \end{array}\right),\nonumber\\
  F^{(3)}_{\rho\mu\nu}(\rho) &=&
                         \left(\begin{array}{ccc}
                        \frac{m_{uu\rho}}{\sinh\!2\rho} & 0 & \frac{m_{u\rho\rho}}{\sinh^2\!2\rho} \\
                        0 & 0 & 0 \\
                        \frac{m_{u\rho\rho}}{\sinh^2\!2\rho} & 0 &  \frac{m_{\rho\rho\rho}}{\sinh^3\!2\rho}
                              \end{array}\right)
 \ea
with
 \ba
 m_{uuu} &=& \{(1+2\mu)(4\mu(1+\mu)-3k^2)-ik(2(1+6\mu+6\mu^2)-k^2)\}\times\nonumber\\
           &&~ \{2(-2+\mu)(12-19\mu+10\mu^2-12k^2)-2ik(44-51\mu+18\mu^2-4k^2)  \nonumber\\
           &&~ +3\mu(29-33\mu+10\mu^2-8k^2-16ik(-2+\mu))\cosh\!2\rho \nonumber\\
           &&~ +6\mu(-2+\mu-ik)(1+2\mu)\cosh\!4\rho
               +\mu(1+\mu)(1+2\mu)\cosh\!6\rho \}
 \ea
 \ba
 m_{uu\rho} &=& \frac{1}{64\cosh^6\!\rho}
                      \{84(1-2\mu)^2(27+90\mu-25\mu^2-100\mu^3+44\mu^4)\nonumber\\
             &&~~~~~~~~~~~~~ +28k^2(115-912\mu-1485\mu^2+4560\mu^3-2172\mu^4) \nonumber\\
             &&~ -12(-6(9-42\mu-523\mu^2+700\mu^3+636\mu^4-1120\mu^5+352\mu^6)\nonumber\\
             &&~~~~~+4k^2(-278+819\mu+1382\mu^2-4340\mu^3+2104\mu^4) \nonumber\\
             &&~~~~~+k^4(-214+2242\mu-2244\mu^2)+40k^6 \nonumber\\
             &&~~~~~+ik(-525-3181\mu+6468\mu^2+7512\mu^3-17080\mu^4+6528\mu^5 \nonumber\\
             &&~~~~~-k^2(477+1791\mu-8822\mu^2+5792\mu^3)+4k^4(-57+116\mu)))\cosh\!2\rho  \nonumber\\
             &&~-3(-6(1-2\mu)^2(108+336\mu-169\mu^2-420\mu^3+220\mu^4)\nonumber\\
             &&~~~~~+2k^2(-516+3296\mu+5585\mu^2-18560\mu^3+9532\mu^4) \nonumber\\
             &&~~~~~-16k^4(29-282\mu+276\mu^2)+64k^6 \nonumber\\
             &&~~~~~+ik(576-6603\mu+13888\mu^2+14768\mu^3-37760\mu^4+15600\mu^5 \nonumber\\
             &&~~~~~-k^2(1168+3739\mu-18272\mu^2+12308\mu^3)+64k^4(-7+13\mu)))\cosh\!4\rho \nonumber\\
             &&~-2(-36+216\mu+2966\mu^2-5400\mu^3-2776\mu^4+8640\mu^5-3520\mu^6 \nonumber\\
             &&~~~~~+4k^2(-236+843\mu+1302\mu^2-4620\mu^3+2664\mu^4) \nonumber\\
             &&~~~~~-2k^4(158-951\mu+942\mu^2)+16k^6 \nonumber\\
             &&~~~~~+ik(-474-2723\mu+7188\mu^2+6152\mu^3-20040\mu^4+9600\mu^5 \nonumber\\
             &&~~~~~-k^2(498+2045\mu-8442\mu^2+6112\mu^3)+24k^4(-7+12\mu)))\cosh\!6\rho \nonumber\\
             &&~+6(-1+2\mu)(2(-9+77\mu^2-46\mu^3-116\mu^4+88\mu^5) \nonumber\\
             &&~~~~~+4k^2(-46+36\mu+358\mu^2-404\mu^3)+4k^4(-4+11\mu) \nonumber\\
             &&~~~~~-ik(-15-175\mu-114\mu^2-468\mu^3+424\mu^4 \nonumber\\
             &&~~~~~+k^2(1+123\mu-190\mu^2)+4k^4))\cosh\!8\rho \nonumber\\
             &&~+6\mu(-1+2\mu)(2\mu(5-10\mu-8\mu^2+16\mu^3)+k^2(4+16\mu-48\mu^2)+2k^4 \nonumber\\
             &&~~~~~+ik(-5+18\mu+28\mu^2-64\mu^3+k^2(-3+16\mu)))\cosh\!10\rho  \nonumber\\
             &&~+(2\mu^2(1-4\mu^2)(1-4\mu^2+3k^2)+ik\mu(1-4\mu^2)(1+12\mu^2-k^2))\cosh\!12\rho \}     \nonumber\\
 \ea
 \ba
 m_{u\rho\rho} &=& \frac{1}{16\cosh^6\!\rho}
                   \{2(12(189+284\mu-518\mu^2-632\mu^3+2037\mu^4-1428\mu^5+308\mu^6)\nonumber\\
             &&~~~~~~~~~~~~~~   -k^2(1010-4695\mu+37881\mu^2-45384\mu^3+15204\mu^4)         \nonumber\\
             &&~~~~~~~~~~~~~~   +k^4(841-2991\mu+2118\mu^2)-80k^6)         \nonumber\\
             &&~~~~~~~~~~~~~~ -2ik(2(2931-2812\mu-5415\mu^2+24814\mu^3-22050\mu^4+5796\mu^5) \nonumber\\
             &&~~~~~~~~~~~~~~ +k^2(210-12953\mu+23325\mu^2-10678\mu^3)+36k^4(-17+25\mu))         \nonumber\\
             &&~-12(630-752\mu+1790\mu^2+2108\mu^3-6848\mu^4+4848\mu^5-1056\mu^6\nonumber\\
             &&~~~~~+k^2(-1167-1619\mu+10438\mu^2-12608\mu^3+4208\mu^4) \nonumber\\
             &&~~~~~+k^4(-433+1643\mu-1122\mu^2)+20k^6 \nonumber\\
             &&~~~~~+ik(-779-1969\mu-3124\mu^2+13816\mu^3-12360\mu^4+3264\mu^5 \nonumber\\
             &&~~~~~+k^2(397-3487\mu+6434\mu^2-2896\mu^3)+8k^4(-21+29\mu)))\cosh\!2\rho \nonumber\\
             &&~-3(-4(360+448\mu-1075\mu^2-1198\mu^3+4033\mu^4-2940\mu^5+660\mu^6) \nonumber\\
             &&~~~~~+k^2(272-3999\mu+23723\mu^2-28792\mu^3+9532\mu^4) \nonumber\\
             &&~~~~~-16k^4(59-221\mu+138\mu^2)+32k^6 \nonumber\\
             &&~~~~~+ik(3536-4486\mu-7450\mu^2+32004\mu^3-29100\mu^4+7800\mu^5 \nonumber\\
             &&~~~~~+k^2(496-7765\mu+14255\mu^2-6154\mu^3)+32k^4(-11+13\mu)))\cosh\!4\rho \nonumber\\
             &&~-2(810-1536\mu+2434\mu^2+2628\mu^3-9728\mu^4+7440\mu^5-1760\mu^6 \nonumber\\
             &&~~~~~+k^2(-949-1677\mu+13818\mu^2-16128\mu^3+5328\mu^4) \nonumber\\
             &&~~~~~+k^4(-599+1653\mu-942\mu^2)+8k^6 \nonumber\\
             &&~~~~~+ik(-1+\mu)(717+2212\mu+6256\mu^2-12600\mu^3+4800\mu^4 \nonumber\\
             &&~~~~~+k^2(-291+4318\mu-3056\mu^2)+144k^4))\cosh\!6\rho \nonumber\\
             &&~+6(-1+\mu)(4(-15-19\mu+35\mu^2+83\mu^3-128\mu^4+44\mu^5) \nonumber\\
             &&~~~~~-k^2(46+321\mu-788\mu^2+404\mu^3)+k^4(-34+44\mu) \nonumber\\
             &&~~~~~-ik(2(-59+101\mu+272\mu^2-518\mu^3+212\mu^4) \nonumber\\
             &&~~~~~+k^2(-66+267\mu-190\mu^2)+4k^4))\cosh\!8\rho \nonumber\\
             &&~+6(1-3\mu+2\mu^2)(2(-3+7\mu+4\mu^2-16\mu^3+8\mu^4)+k^2(-5+28\mu-24\mu^2)+k^4 \nonumber\\
             &&~~~~~+ik(-5-14\mu+52\mu^2-32\mu^3+k^2(-5+8\mu)))\cosh\!10\rho  \nonumber\\
             &&~ + \left(\mu(1-3\mu+2\mu^2)(4\mu(1-3\mu+2\mu^2)+3k^2(1-2\mu)\right.      \nonumber\\
               &&~~~~~~~~~~~~~ +\left.ik(-2+12\mu-12\mu^2+k^2)\right)\cosh\!12\rho\}, \nonumber\\
 \ea
 \ba
 m_{\rho\rho\rho} &=& \frac{1}{16\cosh^6\!\rho}
                   \{4(3(3168+1488\mu+3737\mu^2-13896\mu^3+15848\mu^4-7392\mu^5+1232\mu^6)\nonumber\\
             &&~~~~~~~~~~~~~~   -k^2(9806-30354\mu+74763\mu^2-58848\mu^3+15204\mu^4)         \nonumber\\
             &&~~~~~~~~~~~~~~   +k^4(3362-7782\mu+4236\mu^2)-80k^6)         \nonumber\\
             &&~~~~~~~~~~~~~~ -2ik(32556+26819\mu-125136\mu^2+194416\mu^3-114240\mu^4+23184\mu^5 \nonumber\\
             &&~~~~~~~~~~~~~~+k^2(7920-51493\mu+60576\mu^2-21356\mu^3)+12k^4(-133+150\mu))   \nonumber\\
             &&~+12(6(-1056+144\mu+1003\mu^2-3972\mu^3+4516\mu^4-2112\mu^5+352\mu^6)\nonumber\\
             &&~~~~~-4k^2(-575-4652\mu+1043+\mu^2-8268\mu^3+2104\mu^4) \nonumber\\
             &&~~~~~+2k^4(878-2165\mu+1122\mu^2)+40k^6 \nonumber\\
             &&~~~~~-ik(-7452+5081\mu-36124\mu^2+54952\mu^3-32360\mu^4+6528\mu^5 \nonumber\\
             &&~~~~~+k^2(3696-14033\mu+16914\mu^2-5792\mu^3)+k^4(-444+464\mu)))\cosh\!2\rho \nonumber\\
             &&~+3(6(2139+498\mu+2407\mu^2-9888\mu^3+11224\mu^4-5280\mu^5+880\mu^6) \nonumber\\
             &&~~~~~-2k^2(5561-22802\mu+49529\mu^2-39024\mu^3+9532\mu^4) \nonumber\\
             &&~~~~~+16k^4(247-602\mu+276\mu^2)-64k^6 \nonumber\\
             &&~~~~~-ik(20475+13397\mu-89648\mu^2+133648\mu^3-78640\mu^4+15600\mu^5 \nonumber\\
             &&~~~~~+k^2(6735-32439\mu+38748\mu^2-12308\mu^3)+64k^4(-15+13\mu)))\cosh\!4\rho \nonumber\\
             &&~-2(9486-5892\mu-10870\mu^2+38952\mu^3-44776\mu^4+21120\mu^5-3520\mu^6 \nonumber\\
             &&~~~~~+8k^2(8-3141\mu+7788\mu^2-5754\mu^3+1332\mu^4) \nonumber\\
             &&~~~~~-2k^4(1283-2355\mu+942\mu^2)+16k^6 \nonumber\\
             &&~~~~~+ik(-8913+15535\mu-56124\mu^2+86072\mu^3-49560\mu^4+9600\mu^5 \nonumber\\
             &&~~~~~+k^2(4071-20363\mu+21054\mu^2-6112\mu^3)+24k^4(-17+12\mu)))\cosh\!6\rho \nonumber\\
             &&~+6(-3+2\mu)(2(-114+4\mu-165\mu^2+538\mu^3-396\mu^4+88\mu^5) \nonumber\\
             &&~~~~~+k^2(80-956\mu+1218\mu^2-404\mu^3)+k^4(-52+44\mu) \nonumber\\
             &&~~~~~-ik(-260-205\mu+1714\mu^2-1604\mu^3+424\mu^4 \nonumber\\
             &&~~~~~+k^2(-184+411\mu-190\mu^2)+4k^4))\cosh\!8\rho \nonumber\\
             &&~+6(3-5\mu+2\mu^2)(2(-21+11\mu+50\mu^2-56\mu^3+16\mu^4)-8k^2(5-12\mu+6\mu^2)\nonumber\\
             &&~~~~~+2k^4 +ik(7-130\mu+180\mu^2-64\mu^3+k^2(-17+16\mu)))\cosh\!10\rho  \nonumber\\
             &&~+(3-11\mu+12\mu^2-4\mu^3)(6-22\mu+24\mu^2-8\mu^3-6k^2(1-\mu) \nonumber\\
             &&~~~~~+ik(11-24\mu+12\mu^2-k^2))\cosh\!12\rho\}.
 \ea
From $(\sinh\rho)^{2(\mu-2)}$ in Eq. (51), its asymptotic form is
given by
 \be
 \Phi^{(3)R}_{\rho\rho\rho}\sim e^{2(\mu-2)\rho},
 \ee
which coincides with that of $n=0$ right-moving quasinormal modes
obtained by solving the second-order differential
equation~\cite{DD}. Explicitly, from Eq. (B.32) in Ref.~\cite{DD},
we recover the same asymptotic form of $R_{222}(\xi)\sim
e^{2(\frac{m}{2}-2)\xi}$ for $\xi=\rho$ and $m=2\mu$.


\end{document}